\documentclass[vecphys]{svmult}

\usepackage{makeidx}         
\usepackage{graphicx}        
\usepackage{multicol}        
\usepackage[bottom]{footmisc}

\usepackage{subfigure}
\usepackage{amsmath,amssymb}
\usepackage{graphicx}
\usepackage{rotating}
\usepackage{bm}
\usepackage{color}

\definecolor{blue}{rgb}{0,0,1}

\definecolor{green}{rgb}{0,1,0}

\definecolor{red}{rgb}{1,0,0}

\newcommand{\be}{\begin{eqnarray}}
\newcommand{\ee}{\end{eqnarray}}

\newcommand{\nn}{\nonumber }

\newcommand{\Gk}{\Gamma_k}

\newcommand{\yb}{\bar\psi}
\newcommand{\xsb}{$\chi$SB}
\newcommand{\Nf}{N_{\text{f}}}
\newcommand{\Nc}{N_{\text{c}}}
\newcommand{\pat}{\partial_t}

\def\slash#1{\setbox0=\hbox{$#1$}               
   \dimen0=\wd0                                 
   \setbox1=\hbox{/} \dimen1=\wd1               
   \ifdim\dimen0>\dimen1                        
      \rlap{\hbox to \dimen0{\hfil/\hfil}}      
      #1                                        
   \else                            
      \rlap{\hbox to \dimen1{\hfil$#1$\hfil}}   
      /                                         
   \fi}                                         %

\makeindex             

\begin{document}

\title*{Chiral Phase Boundary of QCD from the Functional Renormalization Group}
\titlerunning{Chiral Phase Boundary of QCD}
\author{Jens Braun}
\institute{Institut f\"ur Theoretische Physik, Philosophenweg 19, 69120 Heidelberg, Germany
\texttt{jbraun@tphys.uni-heidelberg.de}}
%
%
\maketitle

\section{Introduction}

Phase transitions in QCD are currently very actively
researched. Although most of the attention is focused on the phase
transition at finite baryon density and temperature and the expected
critical point in the phase diagram~\cite{Fodor:2001pe,Allton:2003vx},
there are still challenges even at vanishing density, such as a
determination of the order of the phase transition or the flavor
dependence of the chiral phase transtion temperature
\cite{Karsch:2000kv}.  Since lattice QCD simulations have not given
final answers to all these questions so far, complementary
non-perturbative approaches are indispensable to gain a better
understanding of the non-perturbative phenomena.

For our non-perturbative study\footnote{Talk given at the 2006 ECT*
School "Renormalization Group and Effective Field Theory Approaches to
Many-Body Systems", Trento, Italy.} of the chiral phase boundary of
QCD at finite temperature, we use the functional renormalization
group~(RG)~\cite{Wegner:1972ih,Wilson:1973jj,Polchinski:1983gv,Wetterich:1992yh,Morris:1993qb},
applied to a formulation of QCD in terms of microscopic degrees of
freedom given by quarks and gluons.  At low temperature and momentum
scales, QCD can be described well by effective field theories in terms
of ordinary hadronic states. But a hadronic picture is eventually
bound to fail at higher temperature and momentum scales, owing to
asymptotic freedom.  The recent discussion of a strongly interacting
high-temperature phase of QCD suggests that a simple description of
QCD around and above the phase transition temperature does not exist
\cite{Shuryak:2003xe}.  In this case, a first-principles description
in terms of quarks and gluons is most promising to bridge wide ranges
in parameter space.

\addtolength{\textheight}{-1.5cm}

Here we discuss particular problems 
which are accessible in a formulation in terms of quark and 
gluons~\cite{Braun:2005uj,Braun:2006jd}: 
first, we present our results for the running of the gauge coupling at finite temperature.
Second, we discuss the interplay
between gluodynamics and (induced) quark dynamics. 
Finally, we give results for the chiral phase boundary in the plane of 
temperature and number of quark flavors, obtained from such an investigation of the quark-gluon dynamics at finite temperature.

The functional RG yields a flow equation for the effective average
action~$\Gamma_k$~\cite{Wetterich:1992yh},
\begin{equation}
\pat\Gamma_k=\frac{1}{2}\, \text{STr}\, \pat R_k\,
(\Gamma_k^{(2)}+R_k)^{-1}, \quad t=\ln \frac{k}{\Lambda},
\label{eq:flow_eq1}
\end{equation}
where $\Gamma_k$ interpolates between the bare action
$\Gamma_{k=\Lambda}= S$ and the full quantum effective action
$\Gamma=\Gamma_{k=0}$; $\Gamma_{k}^{(2)}$ denotes the second
functional derivative with respect to the fluctuating field. The
regulator function $R_k$ specifies the details of the Wilsonian
momentum-shell integrations, such that the flow of $\Gk$ is dominated 
by fluctuations with momenta $p^2\simeq k^2$. 

It is impossible to study the flow of the most general effective action, 
consisting of all operators that are compatible with the symmetries of 
the theory. Therefore we have to truncate the action to a subset of operators, 
which is not necessarily finite.
Nevertheless, such an approximation of the full theory can also describe reliably
non-perturbative physics, provided the relevant degrees of freedom in the
form of RG relevant operators are kept in the ansatz for the effective
action. This is obviously the most problematic part since 
it requires a lot of physical insight to make the correct physical 
choice.  A first but highly nontrivial check
of a solution to the flow equation is provided by 
a stability analysis of its RG flow, since
insufficient truncations generically exhibit IR instabilities 
of Landau-pole type.

The IR stability of RG flows can be improved by adjusting the
regulator to the spectral flow of $\Gamma^{(2)} _k$ \cite{Gies:2002af,Gies:2003ic,Litim:2002xm}. 
Doing this, we integrate over shells of eigenvalues of $\Gamma^{(2)} _k$ rather than
ordinary canonical momentum shells. 
In a perturbative language, the use of the spectrally adjusted regulator 
allows for a resummation of a larger class of diagrams. Such an improvement has been 
successfully applied to the study of QCD at zero and 
finite temperature~\cite{Gies:2002af,Gies:2003ic,Braun:2006jd} and is the underlying technical 
ingredient for the results presented below.

In the subsequent sections, we use the exponential regulator~\cite{Wetterich:1992yh} of the
form $R_k(\Gamma^{(2)}_k)=\Gamma^{(2)}_k /[\exp (\Gamma^{(2}_k/\mathcal Z_k 
k^2)-1]$, where $\mathcal Z_k$ denotes the wave function
renormalization of the fields. 

\section{Running gauge coupling at finite temperature}

In this section, we present our study of the running strong coupling 
at finite temperature. While an investigation of the running of the coupling  
is interesting in its own right, we will also show in the subsequent sections 
that it represents a key ingredient of our study of the chiral phase boundary.
\begin{figure}[t]
\begin{center}
\subfigure[]{\label{betafct}\includegraphics[%
  clip,
  scale=0.76]{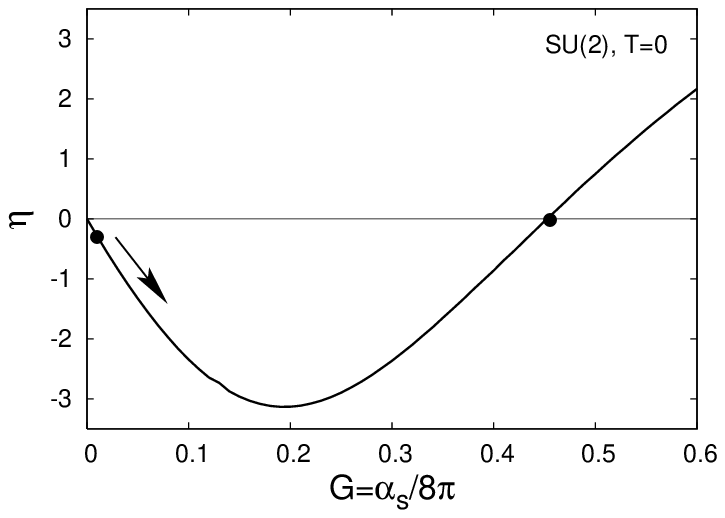}}\hfill
\subfigure[]{\label{coupl_pure_glue}\includegraphics[%
  clip,
  scale=0.46]{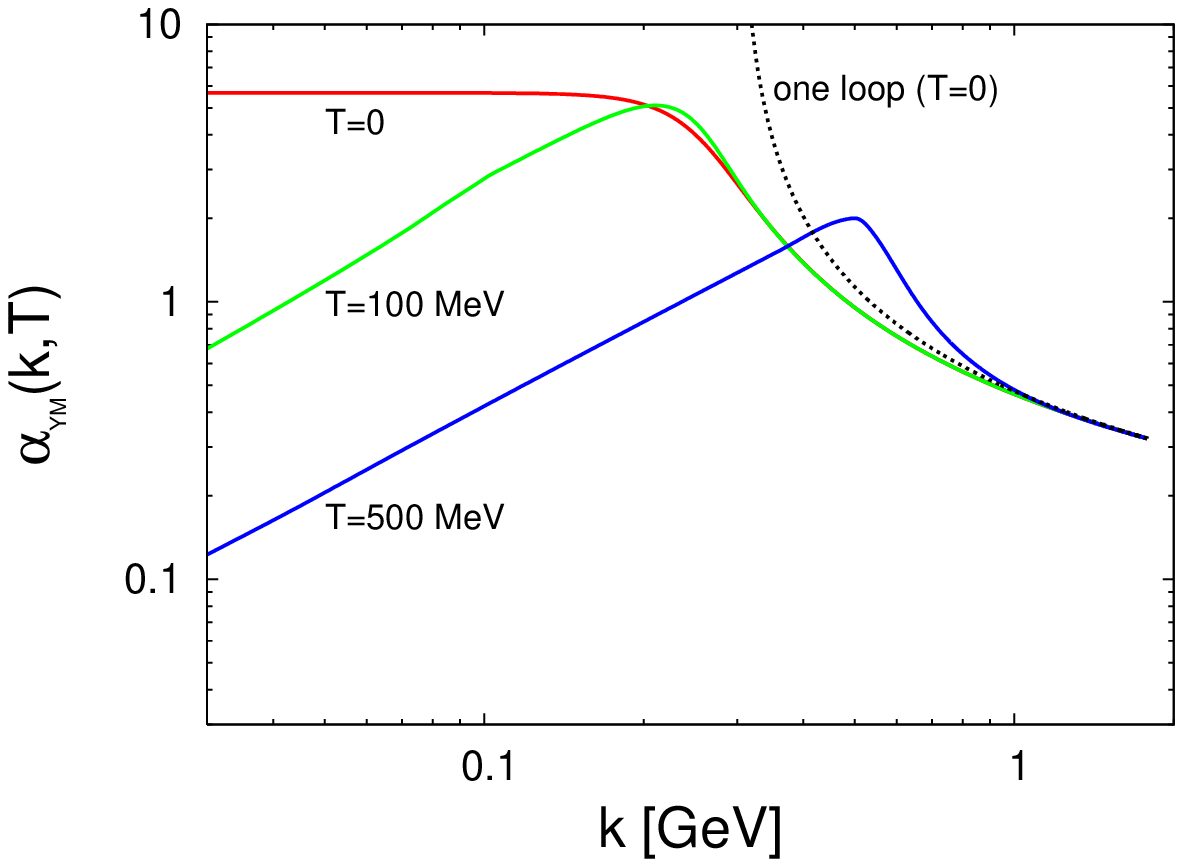}}
\caption{
(a): Anomalous dimension $\eta$ 
as a function of $\alpha/8\pi = g^2/32\pi ^2$ for $SU(2)$ Yang-Mills 
theory at vanishing temperature. 
The anomalous dimension is defined via the background-field wave-function renormalization 
by $\eta=-\ln\partial _t W_k ^{(1)}$. It is related to the flow of the 
coupling as $\pat g^2 _k=\eta g^2 _k$. 
The arrow indicates the RG flow from the initial condition at the UV scale $\Lambda$ (left dot)
to the IR fixed point (right dot). Owing to asymtotic freedom, there exists 
also a fixed point for $\alpha =0$. 
(b):~Running $\mathrm{SU}(3)$ Yang-Mills coupling
$\alpha _{\text{YM}}(k,T)$ as a function of $k$ for temperatures 
$T=0,100,300\,\mathrm{MeV}$, compared to the one-loop 
running for vanishing temperature.}
\label{fig:runcoup}
\end{center}
\end{figure}

Owing to strong coupling, we cannot expect that low-energy gluodynamics are 
reliably described by a small number of gluonic operators. On the contrary, 
infinitely many operators become RG relevant in the low-momentum regime and drive 
the running of the coupling. We truncate the space of possible action
functionals to an {\em infinite} but still tractable set of operators.  
For gauge invariance, we apply the background-field formalism as developed in
\cite{Reuter:1993kw,Freire:2000bq} and follow the strategy of
\cite{Reuter:1997gx,Gies:2002af} for an approximate resolution of the
gauge constraints \cite{Ellwanger:1994iz}. For an introduction to the background-field 
formalism and to background-field flow equations for gauge theories, 
we refer the reader to the lecture notes of H. Gies \cite{LectureHGies}.
Apart from standard gauge-fixing and ghost terms, our truncation of the 
gauge sector consists of an infinite set of operators given by powers of the
Yang-Mills Lagrangian,
\begin{equation}
\Gamma_{k}[A,\bar{\psi},\psi]=\int_x \mathcal W_k(\theta)+\Gamma _k ^{\psi}[A,\bar{\psi},\psi], \quad
\theta=\frac{1}{4} F_{\mu\nu}^a F_{\mu\nu}^a,\label{Wtrunc}
\end{equation}
whereas the quark contributions 
are summarized in $\Gamma _k ^{\psi}$ and will be discussed in Sec. \ref{sec:cpbQCD}. 
For the moment, we restrict our study to the gluonic subspace.
Expanding the function $\mathcal W_k(\theta)=W_k ^{(1)} \theta+ \frac{1}{2}
W_k ^{(2)} \theta^2+\frac{1}{3!} W_k ^{(3)} \theta^3\dots$, the expansion coefficients 
span an infinite set of generalized couplings.  The background-field method \cite{Abbott:1980hw}
provides us with a non-perturbative definition of  the running coupling $g_k$ in terms of 
the background-field wave function renormalization $W_k ^{(1)}$: $g_k ^2 W_k ^{(1)}=\bar{g} ^2$.

Our truncation represents a gradient expansion in the field strength,
which includes arbitrarily
high gluonic correlators projected onto their small-momentum limit and
onto the color and Lorentz structure arising from powers of $F^2$.
A drawback of such gradient expansions is the appearance of
an IR unstable Nielsen-Olesen mode in the spectrum 
\cite{Nielsen:1978rm}. At finite temperature $T$, such a mode will be 
populated by thermal fluctuations, typically 
spoiling perturbative computations, see e. g. Ref.~\cite{Dittrich:1980nh}. 
Our RG approach allows us to resolve this problem with the 
aid of a temperature-dependent regulator which removes the 
unphysical thermal population of the unstable mode. 
Thus, we obtain a strictly positive thermal fluctuation spectrum.
For details of the implementation of such a regulator, we refer to \cite{Braun:2006jd}.

Inserting the truncation~\eqref{Wtrunc} in the flow equation~\eqref{eq:flow_eq1}, we find 
that the running of the coupling is successively driven by all generalized
couplings $W_k ^{(i)}$. Keeping track of all contributions from the flows of
the $W_k ^{(i)}$, we obtain a nonperturbative $\beta_{g^2}$ function in
terms of an infinite asymptotic but resummable series in powers of $g^2$,
\begin{equation}
\beta_{g^2} \equiv \partial _t g^2 _k= \sum_{m=1}^\infty a_m({\textstyle{\frac{T}{k}}},N_c)
\frac{(g^2)^{m+1}}{[2(4\pi)^{2}]^m}. \label{betares}
\end{equation}
The coefficients $a_m$ depend on the temperature $T$ and the rank $N_c$ of 
the gauge group\footnote{Odd powers in the coupling $g$ which arise
in high-temperature small-coupling expansions \cite{Blaizot:1999ap,Blaizot:2000fc} 
are not reproduced in our calculation. This is due to the fact that we have not yet included 
non-local operators in our truncation. In any case, we do not expect that 
these operators are quantitatively important for the determination of the 
chiral phase boundary since, as we will discuss below, chiral symmetry breaking sets in 
at scales $k<T$.}.  
At vanishing temperature, 
the $\beta_{g^2}$ function agrees well with perturbation theory for small
coupling. For larger coupling, we find that the integral representation of
Eq.~\eqref{betares} has a second zero, corresponding to an IR attractive
non-Gau\ss ian fixed point $g^2_\ast>0$, in agreement with the results of 
Ref.~\cite{Gies:2002af}, see Fig.~\ref{betafct}. 
For an explicit representation of the $a_m$'s and
details of the computation, we refer the reader to Ref.~\cite{Braun:2006jd}. Note that the
appearance of an IR fixed point in Yang-Mills theories is a
well-investigated phenomenon in the Landau gauge~\cite{vonSmekal:1997is,Pawlowski:2003hq,Fischer:2004uk}, 
in accordance with the Kugo-Ojima and
Gribov-Zwanziger confinement scenarios~\cite{Kugo:1979gm,Gribov:1977wm}. 
Moreover, it is also
compatible with the existence of a mass gap~\cite{Gies:2002af} in Yang-Mills theory.

As initial condition for the coupling flow,
we use the value of the coupling measured at the
$\tau$ mass scale \cite{Bethke:2004uy}, $\alpha_{\mathrm{s}}=0.322$.  
For scales $k\gg T$, we find agreement with the perturbative running
coupling at zero temperature, as one would naively expect.  
In the IR, the running is strongly modified: The coupling increases towards lower scales until
it develops a maximum near $k\sim T$. Below, the coupling decreases
according to a power law $g^2 \sim k/T$, see Fig.~\ref{coupl_pure_glue}. 
The reason for this behavior can be understood within the 
RG framework: first, the hard gluonic modes decouple from the RG flow at the scale $k\sim T$. 
At this point, the wavelength of fluctuations with momenta $p^2<T^2$ is larger than
the extent of the compactified Euclidean time direction. Hence these
modes become effectively 3-dimensional and their limiting behavior is
governed by the spatial 3$d$ Yang-Mills theory. 
However, the decoupling of the hard modes alone 
cannot explain the decrease of the coupling for scales $k<T$.
The second ingredient which is needed
is the existence of  a non-Gau\ss ian IR fixed point
also in the reduced 3-dimensional theory.
A straightforward matching between the 4$d$ and 3$d$ coupling reveals 
that the observed power law for the 4$d$ coupling is a direct consequence of
the strong-coupling IR behavior in the 3$d$ theory,
$g^2(k\ll T)\approx g^2_{3d,\ast}\, k/T$. Note that 
this asymptotic behavior can be deduced analytically from the 
the integral representation of Eq.~\eqref{betares}, see Ref.~\cite{Braun:2006jd}.
Again, the observed IR behavior at finite temperature is in accordance
with recent results in the Landau gauge~\cite{Maas:2004se}.

\section{Chiral Phase Boundary of QCD}\label{sec:cpbQCD}
For a study of chiral symmetry breaking at finite temperature, 
we extend our truncation~\eqref{Wtrunc}
in two directions. To make this more transparent, we 
divide $\Gamma _k ^{\psi}$ in Eq.~\eqref{Wtrunc}
into two terms, $\Gamma _k ^{\psi}=\Gamma _k ^{\psi-\text{kin.}}+\Gamma _k ^{\psi-\text{int.}}$.
With the first term, we include the quark contributions to all gluonic operators, 
as done in Ref.~\cite{Gies:2004hy} for QED and in Ref.~\cite{Braun:2006jd} for QCD:
\be
\Gamma _k ^{\psi-\text{kin.}} [A,\bar{\psi},\psi]=\int _x \bar{\psi}\slash{D}[A]\psi\,,
\ee
where we confine ourselves to massless quarks. An inclusion of mass terms 
is straightforward and has been done in Refs.~\cite{Braun:2006jd}.
This kinetic term successively contributes to the RG flow of the coupling and 
accounts for the screening nature of fermionic fluctuations.

The second term, $\Gamma _k ^{\psi-\text{int.}}$, accounts for the quark self-interactions.
In a consistent and systematic operator expansion, the lowest non-trivial choice for this 
term is given by\footnote{Qualitatively, we expect our truncation of the fermionic 
self-interactions to be reliable for our purposes, since the feed-back of higher-order 
operators, e. g. $\sim~(\bar{\psi}\psi)^4$, is in general suppressed by the one-loop 
structure of the underlying flow equation.}
\be
\Gamma_k^{\psi-\text{int.}}[\bar{\psi},\psi]&=&\int_x \frac{1}{2} \Big[
  \bar\lambda_-(\text{V--A})+\bar\lambda_+ (\text{V+A}) 
  +\bar\lambda_\sigma (\text{S--P})\nn\\
& &\quad\qquad  +\bar\lambda_{\text{VA}}
  [2(\text{V--A})^{\text{adj}}+({1}/{\Nc})(\text{V--A})] \Big].
\label{equ::truncsym}
\ee
We have classified the  four-fermion interactions 
according to their color and flavor structure:
\be
(\text{V--A})=(\yb\gamma_\mu\psi)^2 + (\yb\gamma_\mu\gamma_5\psi)^2,&\;\;\;&
(\text{V+A}) =(\yb\gamma_\mu\psi)^2 - (\yb\gamma_\mu\gamma_5\psi)^2 ,\nn\\
(\text{S--P})=(\yb^{\chi}\psi^{\xi})^2-(\yb^{\chi}\gamma_5\psi^{\xi})^2,&\;\;\;&
(\text{V--A})^{\text{adj}}=(\yb \gamma_\mu T^a\psi)^2 
   + (\yb\gamma_\mu\gamma_5 T^a\psi)^2.\nn
\ee
Color and flavor singlets are given in the first line.  
Fundamental color ($i,j,\dots$) and flavor ($\chi,\xi,\dots$)
indices are contracted pairwise. The operators with 
non-singlet color or flavor structure are given in the second line, 
where $(\yb^{\chi}\psi^{\xi})^2\equiv \yb^{\chi}\psi^{\xi} 
\yb^{\xi} \psi^{\chi}$, etc., and the $(T^a)_{ij}$ denote the generators of the gauge group in the
fundamental representation.  

In the following, we study the four-fermion couplings in the 
point-like limit, $\lambda _i =\lambda _i (|p_i|\ll k)$. This is
a severe approximation in the chirally broken regime where 
mesons manifest themselves as momentum singularities of the four-fermion 
couplings. Nevertheless, our point-like truncation can be a reasonable approximation in the
chirally symmetric regime\footnote{This has recently been quantitatively
confirmed for the zero-temperature chiral phase transition in
many-flavor QCD \cite{Gies:2005as}.}.
We stress that our truncation~\eqref{equ::truncsym} for the fermionic self-interactions
forms a complete basis, i. e. any
other pointlike four-fermion interaction which is invariant under
\mbox{$\textrm{SU}(\Nc)$} gauge symmetry and 
$\textrm{SU}(\Nf)_{\textrm{L}}\times \textrm{SU}(\Nf)_{\textrm{R}}$
flavor symmetry is reducible by means of Fierz transformations.
Terms accounting for instantons, as U${}_{\text{A}}(1)$-violating
interactions have been neglected as well, since they may become relevant only inside
the \xsb\ regime or for small $\Nf$. 
\begin{figure}[t]
\begin{center}
\subfigure[]{\label{fig:graphs}\includegraphics[%
  clip,
  scale=0.62]{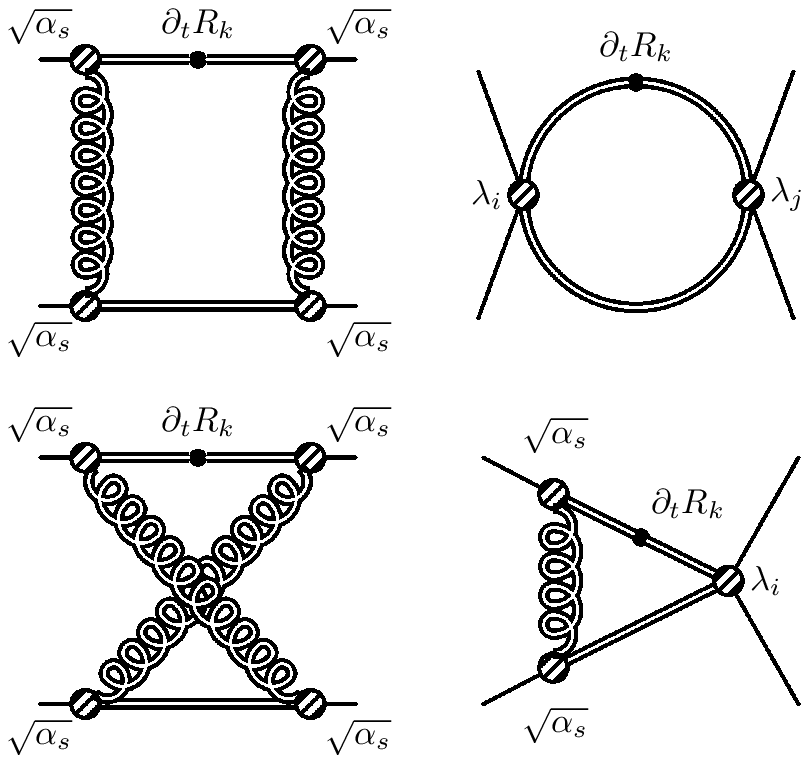}}\hfill
\subfigure[]{\label{fig:parab}\includegraphics[%
  clip,
  scale=0.63]{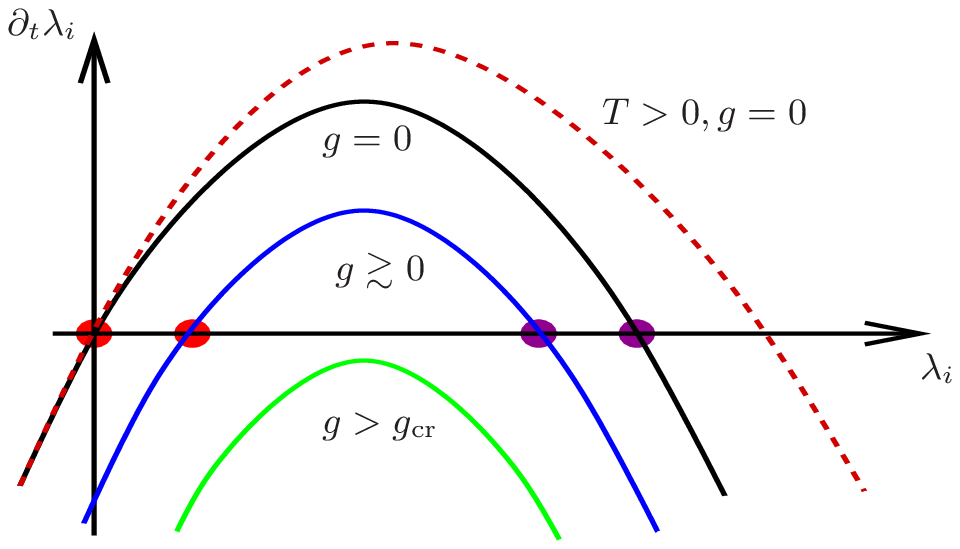}}
\caption{(a): Representation of the one-particle irreducible (1PI)
graphs which are contained in the RG flow equations for the four-fermion 
interactions. The double-lines represent the fully dressed propagator and 
the solid black dots denote the insertion of $\partial _t R_k$ in the loop. The 
four-fermion couplings $\lambda _i$ are defined in Eq.~\eqref{equ::truncsym}. Note that 
diagrams with the same topology but with the regulator insertion attached to the other internal lines 
are present in the RG flow as well.
(b): Sketch of a typical $\beta$ function for the fermionic
  self-interactions $\lambda_i$  at vanishing temperature 
  for zero gauge coupling (upper solid curve), for finite gauge coupling $g>0$ (middle/blue solid curve) 
  and for gauge couplings larger than the critical coupling $g>g_{\text{cr}}$
  (lower/green solid curve).  The effect of finite ratios $T/k$ 
  is illustrated for vanishing gauge coupling by the dashed (red) line.}
\end{center}
\end{figure}

Introducing the dimensionless couplings $\lambda_i =k^2 \bar\lambda _i$, the
$\beta$ functions for the four-fermion couplings can be written in the form
\begin{equation}
\pat\lambda _i =2\lambda _i - \lambda _j  A_{jk} \lambda _k - b_j \lambda _j g^2 - c_i  g^4. 
\label{eqlambda}
\end{equation}
Here we use a straightforward finite-temperature
generalization of the flow equations for the $\bar\lambda_i$, which have been 
derived and analyzed in \cite{Gies:2003dp,Gies:2005as}. 
The quantities $A$, $b$ and $c$ are dependent on the ratio $T/k$, where 
$A$ is a matrix, and $b$ and $c$ are vectors in the space of $\lambda$ couplings;
see Ref. \cite{Braun:2006jd} for an explicit representation of the flow equations.
The various terms arising on the RHS of the flow equation of the $\bar{\lambda}_i$s 
can be related straightforwardly to one-particle irreducible Feynman-graphs, which are depicted in 
Fig.~\ref{fig:graphs}.
As initial conditions for the four-fermion couplings, we use $\bar\lambda_i\to 0$ for
$k\to\Lambda\to\infty$. This choice ensures that the $\bar{\lambda}_i$ 
are generated solely by quark-gluon dynamics from first principles.
This point is very important, as it is contrary to, e.g., the Nambu--Jona-Lasinio model, where the
four-fermion couplings serve as independent input parameters.

Our truncation provides a simple picture for the chiral dynamics illustrated 
by Fig.~\ref{fig:parab}: At vanishing gauge coupling, we find fixed points 
for the four-fermion couplings at $\lambda _i \neq 0$ and $\lambda _i =0$, 
where the latter ones are IR attractive. By increasing the gauge coupling $g$, 
the RG flow generates quark self-interactions of order $\lambda\sim g^4$, 
resulting in a shift of the fixed points. 
Moreover, we observe that the four-fermion couplings 
approach fixed points $\lambda_\ast$ in the IR, if the gauge coupling 
in the IR remains smaller than a critical value $g_{\text{cr}}$.
At these fixed points, QCD remains in the chirally invariant phase.
Increasing the gauge coupling beyond the critical coupling $g>g_{\text{cr}}$, 
the gauge-fluctuation induced $\lambda$'s become strong enough to 
contribute as relevant operators to the RG flow. This is reflected 
in a destabilization of the IR fixed points \cite{Gies:2003dp,Gies:2005as}.
In this case, the four-fermion couplings increase rapidly and approach a divergence 
at a finite scale~$k=k_{\text{\xsb}}$. Indeed, this strong increase indicates the 
formation of chiral quark condensates and therefore the onset of chiral symmetry breaking.
We recall the NJL model as an illustration: There, the mass parameter $m^2$ of the bosonic fields 
in the partially bosonized action is inversely proportional to the 
four-fermion coupling, $\bar{\lambda}\sim 1/m^2$. In addition, we know from such NJL-type models 
that their effective action in the bosonized form corresponds to a Ginzburg-Landau effective potential for the order parameter. In its simplest form, the order parameter is given by the expectation value of a scalar field.  
Symmetry breaking is then reflected in a non-trivial minimum of the this potential.
Consequently, the scale $k_{\text{\xsb}}$ at which the four-fermion couplings 
diverge is a good measure for the chiral  symmetry breaking scale. At this scale, 
the effective potential for the order parameter becomes flat and starts to develop a 
nonzero vacuum expectation value.
\begin{figure}[!t]
\begin{center}
\includegraphics[%
  clip,
  scale=0.85]{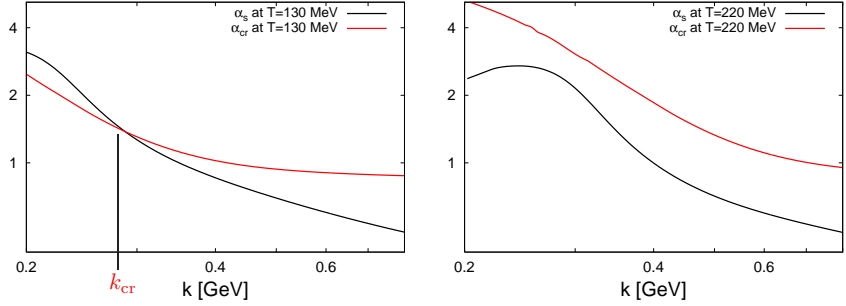}
\caption{\label{alpha_alpha_c} The figures show the 
running QCD coupling
$\alpha_{\mathrm{s}} (k,T)$ for $\Nf=3$ massless quark flavors and
$\Nc=3$ colors and the critical value of the running coupling
$\alpha_{\mathrm{cr}} (k,T)$ as a function of $k$ for two values 
of the temperature, $T=130\,\mathrm{MeV}$ (left panel) and $T=220\,\mathrm{MeV}$ (right
panel). The existence of the
$(\alpha_{\mathrm{s}},\alpha_{\mathrm{cr}})$ intersection point at the scale $k_{\text{cr}}$ 
in the left panel indicates that the quark dynamics can become critical.}
\end{center}
\end{figure}

At this point, we have traced the question of the onset of chiral symmetry 
breaking back to the strength of the coupling $g$ relative to the critical coupling $g_{\text{cr}}$.
At zero temperature, the critical coupling for three colors and not 
too many (massless) quark flavors is much smaller than the IR fixed point 
value of the coupling. Therefore QCD exhibits broken chiral symmetry 
at zero temperature.

At finite temperature, the running of the gauge coupling is
strongly modified in the IR. Moreover, the critical coupling becomes larger 
for increasing ratios~$T/k$. This reflects that the formation of 
quark condensates require stronger interactions at finite temperatures, since 
the quarks become stiffer due to the thermal masses.

In Fig. \ref{alpha_alpha_c}, we show the running coupling
$\alpha_{\mathrm{s}}$ and its critical value $\alpha_{\mathrm{cr}}$
for $T=130\,\mathrm{MeV}$ and $T=220\,\mathrm{MeV}$ as a function of
the regulator scale $k$. The intersection point $k_{\text{cr}}$
between both gives the scale where the chiral quark dynamics become
critical. Above this scale, the system is in the chiral symmetric regime, below 
it quickly runs into the broken regime.
The critical temperature can now be estimated using the lowest temperature 
for which no intersection point between 
$\alpha_{\mathrm{s}}$ and $\alpha_{\mathrm{cr}}$ exists.
We find $T_{\mathrm{cr}}\approx
186\,\mathrm{MeV}$ for $\Nf=2$ and $T_{\mathrm{cr}}\approx
161\,\mathrm{MeV}$ for $\Nf=3$ massless quark flavors in good
agreement with lattice simulations \cite{Karsch:2000kv}.
We stress that no other parameter except for the running coupling at the 
$\tau$ mass scale has been used as input for the calculation of the 
critical temperatures.

\begin{figure}[!t]
\begin{center}
\subfigure[]{\label{tc_nf}\includegraphics[%
  clip,
  scale=0.45]{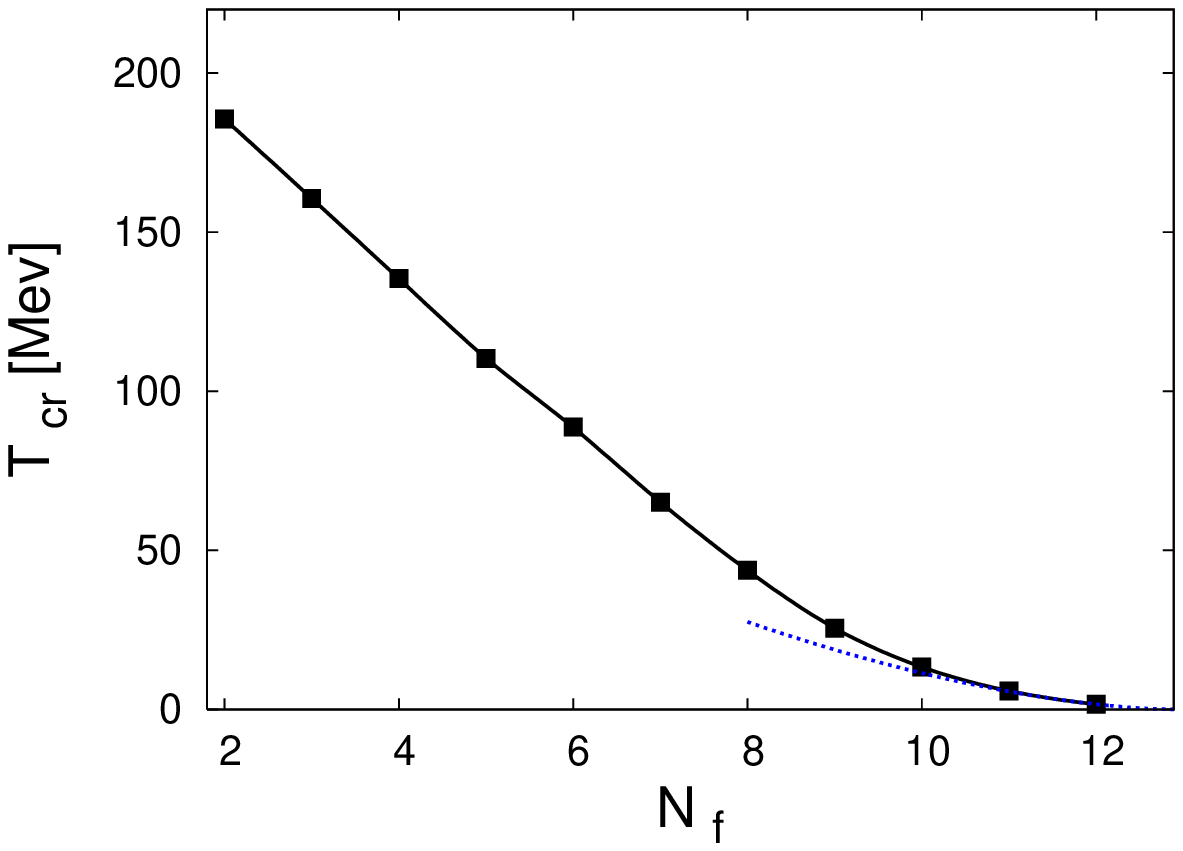}}\hfill
\subfigure[]{\label{crit_exp}\includegraphics[%
  clip,
  scale=0.45]{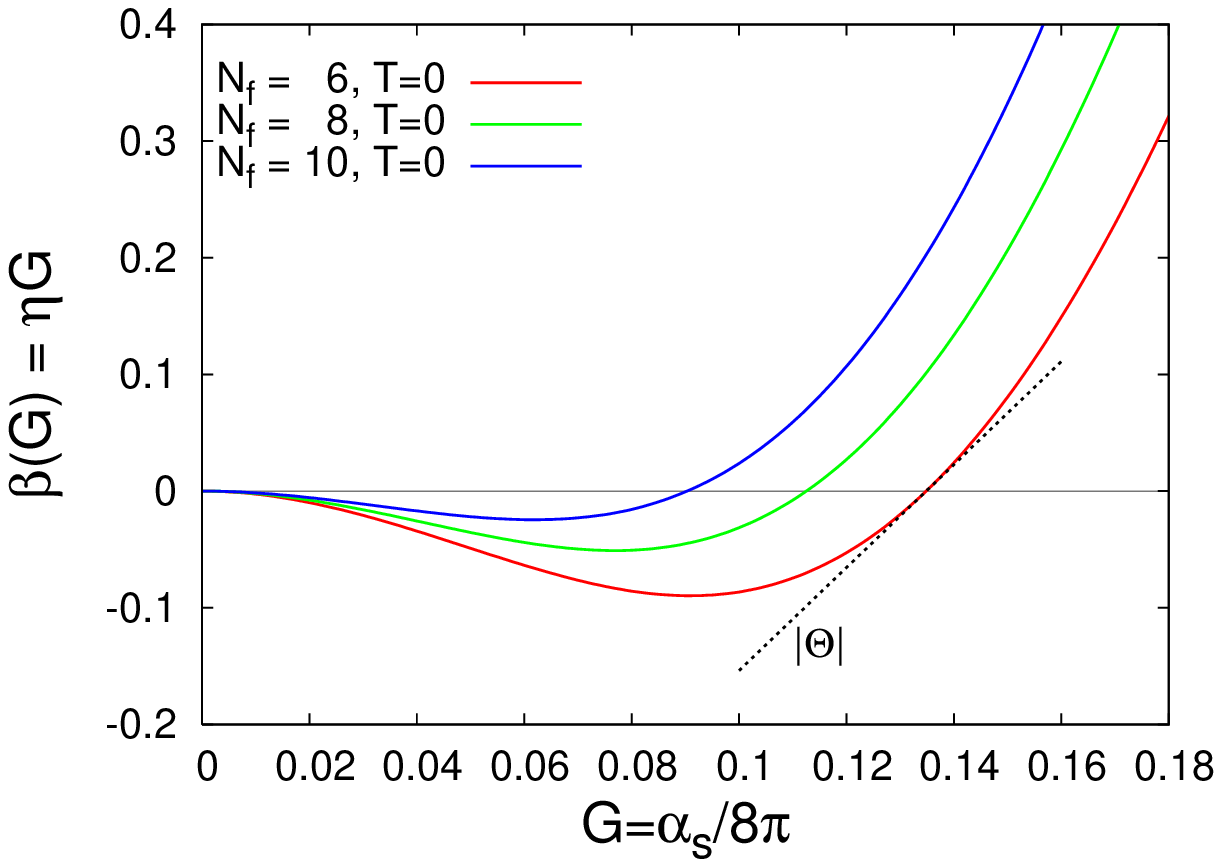}}
\caption{(a): Chiral phase-transition temperature
$T_{\text{cr}}$ versus the number of massless quark flavors $\Nf$. 
The flattening at $\Nf\gtrsim10$ is a consequence of the IR fixed-point structure, see text. 
The analytic prediction~\eqref{eq:TcrTheta} for the phase boundary is depicted by the dashed (blue) line.
(b): The figure shows the $\beta$ function for $\Nf=6,8,10$ massless quark flavors (from bottom to top). 
For illustration, we also show the linear approximation of the $\beta$ function for $\Nf =6$ whose slope 
gives the corresponding critical exponent $\Theta$.}
\end{center}
\end{figure}

Finally, we discuss the chiral phase boundary in the plane of the temperature 
and number $\Nf$ of massless quark flavors computed within our approach, 
see Fig. \ref{tc_nf}. For small $\Nf$, we observe an almost linear decrease of the critical temperature 
for increasing $\Nf$ with a slope of $\Delta T_{\mathrm{cr}}=T(\Nf)-T(\Nf+1)\approx
25\,\mathrm{MeV}$. Additionally we find the existence of 
a critical number of quark flavors, $\Nf ^{\mathrm{cr}}=12$, above which no chiral phase
transition occurs. Since $\Nf ^{\mathrm{cr}}<\Nf ^{\mathrm{a.f.}}=\frac{11}{2}
N_{c}=16.5$, our study confirms the existence of a regime where QCD is
chiral symmetric but still asymptotically free.
Our value for $\Nf^{\mathrm{cr}}$ is in agreement with other studies based on the 2-loop $\beta$-function~\cite{Banks:1981nn,Miransky:1996pd,Appelquist:1996dq}. A remarkable 
feature of the phase boundary is its shape near $\Nf^{\text{cr}}$. It is a generic prediction of the
IR fixed-point scenario and can be understood from analytic arguments~\cite{Braun:2006jd}:
For $\Nf \approx \Nf^{\text{cr}}$, where fermionic screening of color charges keeps the 
coupling small, the coupling must almost reach its maximal value in order 
to drive the quark sector to criticality. This maximal value can be roughly estimated from 
the fixed point value $g ^2 _\ast$ of the coupling at zero temperature. In the fixed point 
regime, the $\beta_{g^2}$ function is essentially given by the linear term of an expansion 
around the fixed point, $\beta_{g^2}=-\Theta\, (g^2-g_\ast^2)$, where $\Theta$ denotes 
the {\em universal} critical exponent. 
For illustration, we show the $\beta_{g^2}$ functions 
for $\Nf=6,8,10$ at zero temperature in Fig.~\ref{crit_exp}.
Using the condition $g^2(k_{\text{cr}})=g^2_{\text{cr}}$, 
we can relate the solution of the $\beta_{g^2}$ function near the fixed 
point to the scale~$k_{\text{cr}}$, where the coupling $g^2$ exceeds its critical value 
$g ^2 _{\text{cr}}$. The fact that $k_{\text{cr}}$ sets  
the scale for the critical temperature $T_{\text{cr}}$ finally yields the following 
relation for the critical temperature near $\Nf^{\text{cr}}$~\cite{Braun:2006jd}:
\begin{equation}
T_{\text{cr}}\sim k_0 |\Nf -\Nf^{\text{cr}}|^{-\frac{1}{\Theta}},
\label{eq:TcrTheta}
\end{equation}
where $\Theta$ should be evaluated at $\Nf^{\text{cr}}$ and the scale $k_0$ is 
implicitly defined by a suitable definition for the $\beta_{g^2}$~function.
Relation~\eqref{eq:TcrTheta} is remarkable since it relates 
two {\em universal} quantities to each other: the shape 
of the phase boundary and the IR critical exponent $\Theta$. The good agreement
between the full numerical result for the phase boundary and the 
analytic result Eq.~\eqref{eq:TcrTheta} around $\Nf ^{\text{cr}}$ is shown 
in Fig.~\ref{tc_nf}.
This allows us to conclude that the symmetry status of the system for a large number of 
quark flavors is governed by the fixed-point regime where dimensionful scales such as
$\Lambda_{\text{QCD}}$ lose their importance. 
 \section{Summary and Outlook}
We have shown that the functional RG allows for a systematic and 
consistent expansion of QCD. Our truncation of the effective action 
based on an operator 
expansion seems to be promising for a study of QCD at zero and 
finite temperature, at least in the chirally symmetric regime.
We have determined the chiral phase boundary of QCD in the
plane of temperature and flavor number. Our quantitative results 
for small $\Nf$ are in agreement 
with lattice simulations for $\Nf=2,3$. For large $\Nf$ we
observe a flattening of the phase boundary near 
$\Nf^{\text{cr}}$, owing to the IR fixed-point structure of QCD.  
Future extensions should include mesonic operators which can
be treated with RG rebosonization techniques \cite{Gies:2001nw,Gies:2002hq}.
We refer the reader to the lecture notes of H.~Gies~\cite{LectureHGies} for an introduction 
to such techniques. An extension of our present approach in this direction 
would not only provide access to the broken phase and mesonic
properties, but also permit a study of the order of the phase
transition. 

The author would like to thank H. Gies for his collaboration on the
work presented here, and is grateful to him and to B. Klein for
proofreading the manuscript.  The author acknowledges support from the
GSI Darmstadt, from the DFG under contract Gi~328/1-3 (Emmy-Noether
program) and from a fellowship provided by the organizers of the ECT*
school.


\begin{thebibliography}{99.}

\bibitem{Fodor:2001pe}
  Z.~Fodor and S.~D.~Katz:
  JHEP {\bf 0203}, 014 (2002)

\bibitem{Allton:2003vx}
  C.~R.~Allton, S.~Ejiri, S.~J.~Hands, O.~Kaczmarek, F.~Karsch, E.~Laermann and C.~Schmidt:
  Phys.\ Rev.\ D {\bf 68}, 014507 (2003)

\bibitem{Karsch:2000kv}
F.~Karsch, E.~Laermann and A.~Peikert:
Nucl.\ Phys.\ B {\bf 605} (2001) 579 
%

\bibitem{Wegner:1972ih}
  F.~J.~Wegner and A.~Houghton:
  Phys.\ Rev.\ A {\bf 8}, 401 (1973)

\bibitem{Wilson:1973jj}
  K.~G.~Wilson and J.~B.~Kogut:
  Phys.\ Rept.\  {\bf 12}, 75 (1974)

\bibitem{Polchinski:1983gv}
  J.~Polchinski:
  Nucl.\ Phys.\ B {\bf 231}, 269 (1984)
 
\bibitem{Wetterich:1992yh}
C.~Wetterich: 
Phys.\ Lett.\ B {\bf 301} (1993) 90

\bibitem{Morris:1993qb}
  T.~R.~Morris:
  Int.\ J.\ Mod.\ Phys.\ A {\bf 9}, 2411 (1994)

\bibitem{Shuryak:2003xe}
E.~Shuryak:
Prog.\ Part.\ Nucl.\ Phys.\  {\bf 53}, 273 (2004)

\bibitem{Braun:2005uj}
  J.~Braun and H.~Gies:
  arXiv:hep-ph/0512085

\bibitem{Braun:2006jd}
  J.~Braun and H.~Gies:
  JHEP {\bf 0606}, 024 (2006)

\bibitem{Gies:2002af}
H.~Gies:
Phys.\ Rev.\ D {\bf 66}, 025006 (2002) 

\bibitem{Gies:2003ic}
H.~Gies:
Phys.\ Rev.\ D {\bf 68}, 085015 (2003)

\bibitem{Litim:2002xm}
D.~F.~Litim and J.~M.~Pawlowski:
Phys.\ Rev.\ D {\bf 66}, 025030 (2002)

\bibitem{Reuter:1993kw} 
M.~Reuter and C.~Wetterich:
Nucl.\ Phys.\ B {\bf 417}, 181 (1994)
%

\bibitem{Freire:2000bq}
F.~Freire, D.~F.~Litim and J.~M.~Pawlowski:
Phys.\ Lett.\ B {\bf 495}, 256 (2000)

\bibitem{Reuter:1997gx}
M.~Reuter and C.~Wetterich:
Phys.\ Rev.\ D {\bf 56}, 7893 (1997)

\bibitem{Ellwanger:1994iz}
U.~Ellwanger:
Phys.\ Lett.\ B {\bf 335} (1994) 364

\bibitem{LectureHGies} H. Gies: 
arXiv: hep-ph/0611146, 
contribution to this volume (2006).

\bibitem{Abbott:1980hw}
L.~F.~Abbott:
Nucl.\ Phys.\ B {\bf 185}, 189 (1981)

\bibitem{Nielsen:1978rm}
N.~K.~Nielsen and P.~Olesen:
Nucl.\ Phys.\ B {\bf 144}, 376 (1978)

\bibitem{Dittrich:1980nh}
W.~Dittrich and V.~Schanbacher:
Phys.\ Lett.\ B {\bf 100}, 415 (1981) 

\bibitem{Blaizot:1999ap}
J.~P.~Blaizot, E.~Iancu and A.~Rebhan:
Phys.\ Lett.\ B {\bf 470}, 181 (1999)

\bibitem{Blaizot:2000fc}
 J.~P.~Blaizot, E.~Iancu and A.~Rebhan:
Phys.\ Rev.\ D {\bf 63}, 065003 (2001)

\bibitem{vonSmekal:1997is}
L.~von Smekal, R.~Alkofer and A.~Hauck:
Phys.\ Rev.\ Lett.\  {\bf 79}, 3591 (1997)

%
\bibitem{Pawlowski:2003hq}
J.~M.~Pawlowski, D.~F.~Litim, S.~Nedelko and L.~von Smekal:
Phys.\ Rev.\ Lett.\  {\bf 93}, 152002 (2004)
%

\bibitem{Fischer:2004uk}
C.~S.~Fischer and H.~Gies:
JHEP {\bf 0410}, 048 (2004)

%
\bibitem{Kugo:1979gm}
T.~Kugo and I.~Ojima: 
  { Prog. Theor. Phys. Suppl.} {\bf 66} (1979) 1 

\bibitem{Gribov:1977wm}
V.~N.~Gribov:
Nucl.\ Phys.\ B {\bf 139}, 1 (1978) 

\bibitem{Bethke:2004uy}
  S.~Bethke:
  Nucl.\ Phys.\ Proc.\ Suppl.\  {\bf 135} (2004) 345

\bibitem{Maas:2004se}
A.~Maas, J.~Wambach, B.~Gruter and R.~Alkofer:
Eur.\ Phys.\ J.\ C {\bf 37}, 335 (2004)

\bibitem{Gies:2004hy}
H.~Gies and J.~Jaeckel:
Phys.\ Rev.\ Lett.\  {\bf 93}, 110405 (2004)

\bibitem{Gies:2005as}
  H.~Gies and J.~Jaeckel:
  Eur.\ Phys.\ J.\ C {\bf 46}, 433 (2006)

\bibitem{Gies:2003dp}
H.~Gies, J.~Jaeckel and C.~Wetterich:
Phys.\ Rev.\ D {\bf 69} (2004) 105008

\bibitem{Banks:1981nn}
T.~Banks and A.~Zaks:
Nucl.\ Phys.\ B {\bf 196}, 189 (1982) 
%

\bibitem{Miransky:1996pd}
V.~A.~Miransky and K.~Yamawaki:
Phys.\ Rev.\ D {\bf 55}, 5051 (1997)
%

\bibitem{Appelquist:1996dq}
T.~Appelquist, J.~Terning and L.~C.~R.~Wijewardhana:
Phys.\ Rev.\ Lett.\  {\bf 77}, 1214 (1996)

\bibitem{Gies:2001nw}
  H.~Gies and C.~Wetterich:
  Phys.\ Rev.\ D {\bf 65} (2002) 065001

\bibitem{Gies:2002hq}
H.~Gies and C.~Wetterich:
Phys.\ Rev.\ D {\bf 69}, 025001 (2004)
%

\end{thebibliography}

\end{document}